\documentclass[10pt, twocolumn, pre, aps, superscriptaddress, showpacs]{revtex4-1}
\usepackage{amsmath, graphicx, subfigure, tikz}
\begin{document}

\title{Driven and Non-Driven Surface Chaos in Spin-Glass Sponges}
\author{Yiğit Ertaç Pektaş}
    \affiliation{Department of Physics, Bo\u{g}azi\c{c}i University, Bebek, Istanbul 34342, Turkey}
\author{E. Can Artun}
    \affiliation{T\"UBITAK Research Institute for Fundamental Sciences, Gebze, Kocaeli 41470, Turkey}
    \affiliation{Faculty of Engineering and Natural Sciences, Kadir Has University, Cibali, Istanbul 34083, Turkey}
\author{A. Nihat Berker}
    \affiliation{Faculty of Engineering and Natural Sciences, Kadir Has University, Cibali, Istanbul 34083, Turkey}
    \affiliation{T\"UBITAK Research Institute for Fundamental Sciences, Gebze, Kocaeli 41470, Turkey}
    \affiliation{Department of Physics, Massachusetts Institute of Technology, Cambridge, Massachusetts 02139, USA}

\begin{abstract}
A spin-glass system with a smooth or fractal outer surface is studied by renormalization-group theory, in bulk spatial dimension $d=3$.  Independently varying the  surface and bulk random-interaction strengths, phase diagrams are calculated.  The smooth surface does not have spin-glass ordering in the absence of bulk spin-glass ordering and always has spin-glass ordering when the bulk is spin-glass ordered.  With fractal ($d > 2$) surfaces, a sponge is obtained and has surface spin-glass ordering also in the absence of bulk spin-glass ordering.  The phase diagram has the only-surface-spin-glass ordered phase, the bulk and surface spin-glass ordered phase, and the disordered phase, and a special multicritical point where these three phases meet.  All spin-glass phases have distinct chaotic renormalization-group trajectories, with distinct Lyapunov and runaway exponents which we have calculated.

\end{abstract}
\maketitle

\section{The Spin-Glass Sponge}

Spin-glass systems have an inherent quantifiable chaos under scale change \cite{McKayChaos,McKayChaos2,BerkerMcKay,McKayChaos4} and thus provide a universal classification and clustering scheme for complex phenomena \cite{classif}.   However, all materials are largely observed from their surfaces, so that we have studied in $d=3$ the spin-glass systems with a surface.  We find both bulk-chaos-driven surface chaos and surface chaos in the absence of bulk chaos.  The latter phenomenon does occur not on flat surfaces, but on convoluted fractal surfaces, namely in spin-glass sponges.  All chaoses are numerically observed and quantified by distinctive Lyapunov exponents.  These results may further refine the spin-glass chaos classification scheme \cite{classif}, in systems with identifiable peripheral complexity.

\begin{figure}[ht!]
\centering
\includegraphics[scale=0.5]{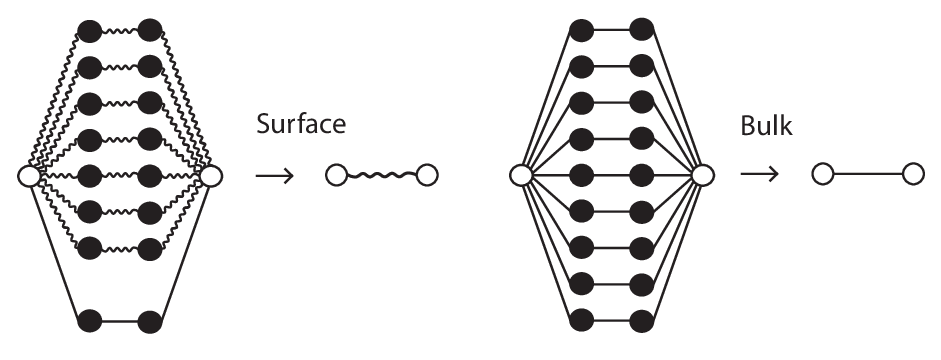}
\caption{Exact renormalization-group transformation for the surfaced hierarchical model.  The straight and squiggly lines represent bulk and surface bonds, respectively.  The hierarchical lattice is constructed by proceeding in the opposite direction, multiply self-imbedding and cross-imbedding the double graphs.  Such a double-graph hierarchical model construction was previously used in the study of anisotropic spin, percolation, high-$T_C$ superconductivity, and spin-glass systems \cite{AnisoModels,AnisoSuperconduct,AnisoSpinGlass}.}
\end{figure}

\begin{figure}[ht!]
\centering
\includegraphics[scale=0.45]{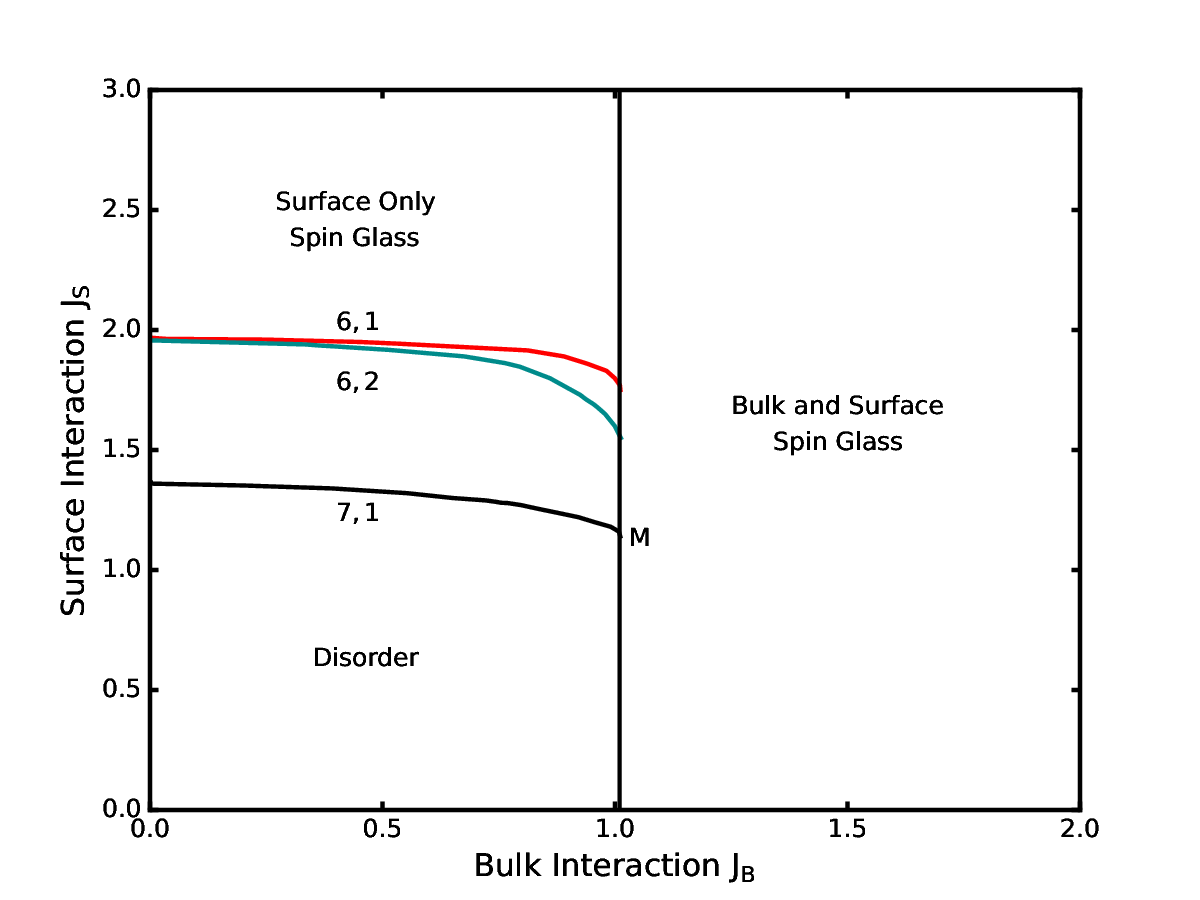}
\caption{Calculated phase diagrams for sponge surface dimensionalities $d_S = 2.77$ and 2.63 are shown.  The bulk contributions to the surface are $n=1$ and 2. The sponges are indicated with $(b^{d_S},n)$ next to the surface-only chaos boundary. These spin-glass sponges exhibit surface spin-glass ordering in the absence of bulk spin-glass ordering.  The phase diagram has a surface-only spin-glass ordered phase, a surface-and-bulk spin-glass ordered phase, and a disordered phase.  The bulk ordering is not affected by the surface interaction, so that the phase boundary to the surface-and-bulk spin-glass ordered phase is a vertical line in the $J_SJ_B$ phase diagram, covering the phase transitions to both the surface-only spin-glass ordered phase and the disordered phase.  The bulk interaction $J_B$ enhances surface-only spin-glass ordering, even in the absence of bulk spin-glass ordering, as seen from the downwards deviation, as the bulk interaction $J_B$ increases, of the near-horizontal phase boundary. On the other hand, as expected, the two phase boundaries for the same $d_S$ merge as $J_B$ goes to zero. The three phase boundary lines meet at the special multicritical point M (marked only for $(7,1)$).  Surface and bulk interaction distributions of 15000 systems each are used for these phase diagram calculations.}
\end{figure}

\begin{figure}[ht!]
\centering
\includegraphics[scale=0.37]{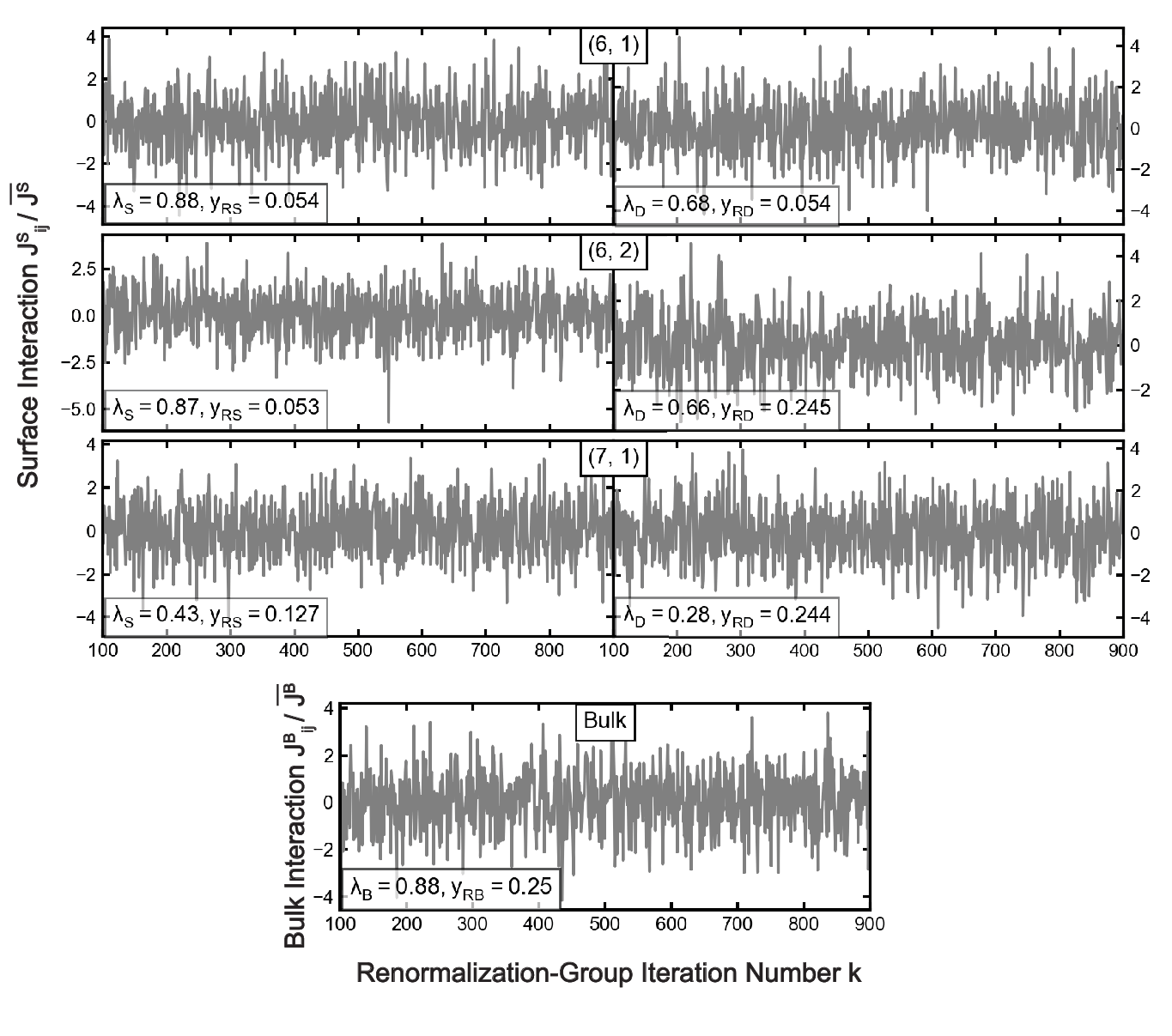}
\caption{For each of our three spin-glass sponges $(7,1),(6,2),(6,1)$, the chaotic renormalization-group trajectories for sole surface chaos $(S)$, driven surface chaos $(D)$, and bulk chaos $(B)$.  The calculated Lyapunov exponents for sole surface chaos $\lambda_S$, driven surface chaos $\lambda_D$, and bulk chaos $\lambda_B$ are also given.  The calculated runaway exponents for sole surface chaos $y_{RS}$, driven surface chaos $y_{RD}$, and bulk chaos $y_{RB}$ are also given. The bulk is not influenced by the surface and thus the bulk items are the same for all three systems. Surface and bulk interaction distributions of 500 systems each are used for these exponent calculations.}
\end{figure}

The surfaced spin-glass system is defined by the Hamiltonian
\begin{equation}
-\beta \mathcal{H}=\sum_{\langle ij \rangle}^S J_{ij}^S s_i s_j + \sum_{\langle ij \rangle}^B J_{ij}^B s_i s_j \,,
\end{equation}
where $\beta=1/kT$, $s_i = \pm1$ at each site $i$ of the semi-infinite $d$-dimensional lattice, and the first (second) sum $\langle ij \rangle$ is over all pairs of nearest-neighbor sites in the surface (bulk). The surface bond $J_{ij}^S$ is ferromagnetic $+J_S>0$ or antiferromagnetic $-J_S$ with probabilities $1-p_S$ and $p_S$, respectively. The bulk bond $J_{ij}^B$ is ferromagnetic $+J_B>0$ or antiferromagnetic $-J_B$ with probabilities $1-p_B$ and $p_B$, respectively.  We concentrate on full realization of spin-glass order, and therefore study $p_S=p_B=0.5$.  The complete range of random interaction strengths $J_S$ and $J_B$ is studied and phase diagrams are obtained.

\section{Renormalization-Group Theory of Surfaced Systems}

The renormalization-group theory of surfaced systems is most easily done with the Migdal-Kadanoff procedure \cite{Migdal,Kadanoff} or, equivalently, the exactly solvable hierarchical models \cite{BerkerOstlund,Kaufman1,Kaufman2,BerkerMcKay}.  The procedure is similar to the solution of anisotropic spin, percolation, high-$T_C$ superconductivity, spin-glass systems, where two interconnected graphs are used \cite{AnisoModels,AnisoSuperconduct,AnisoSpinGlass}.

The Migdal-Kadanoff procedure is a physically intuited approximation:  A system is rendered renormalizable by removing some of the bonds.  Thus a scale change can be achieved by decimation.  Then the effect of the removed bonds is added and the renormalized interactions are obtained.  In surfaced systems, the bulk is transformed as just explained, the final addition involving the sum of $b^{d-1}$ decimated bulk bonds, where $d$ is the bulk spatial dimensionality and $b$ is the length-rescaling factor.  For the surface, the final addition involves the sum of $b^{d_S-1}$ decimated surface bonds and $n$ decimated bulk bonds, where $d_S$ is the surface spatial dimensionality and, from the interior, $n=(b^{d-1}-b^{d_S-1})/2$.  For flat surfaces, $d_S=d-1$ and, for fractal surfaces, $d-1<d_S<d$.  Furthermore, we shall use different values of $n$, as long as $n+b^{d_S-1}<b^{d-1}$, thus conserving contact with the outside.

The double-graphical \cite{AnisoModels,AnisoSuperconduct,AnisoSpinGlass} representation of the equivalent, exactly solved hierarchical lattice \cite{BerkerOstlund,Kaufman1,Kaufman2,BerkerMcKay} is shown in Fig. 1.  Exactly solved hierarchical lattices are widely used for a large variety of systems. \cite{Clark,Kotorowicz,ZhangQiao,Jiang,Chio,Myshlyavtsev,Derevyagin,Shrock,Monthus,Sariyer,classif}  Thus, the Migdal-Kadanoff approximation a physically realizable approximation, as is used in turbulence \cite{Kraichnan}, polymers \cite{Flory,Kaufman}, electronic systems \cite{Lloyd}, and therefore a robust approximation (e.g., it will never yield a negative entropy).

Thus, for quenched random systems, a distribution of bulk interactions and a distribution of surface interactions are used in the calculation.  We use, for each distribution, 15000 interactions originally bimodally chosen as described above.  At each renormalization-group step, $b^d$ bulk interactions, randomly chosen from the bulk distribution, give a renormalized bulk interaction; $b^{d_S}$ surface interactions and $bn$ bulk interactions, randomly chosen from their respective distributions, give a renormalized surface interaction.  Both distributions are renormalized (and of course do not remain bimodal) by repeating this operation 15000 times.  Under repeated renormalization-group transformations, each phase, phase boundary and, in this case, special multicritical point is attracted to its own fixed distribution and, thus, phase diagrams are obtained.

\section{Phase Diagrams of the Spin-Glass Sponge}

Calculated phase diagrams for sponge surface dimensionalities $d_S = 2.77$ and 2.63 are shown in Fig. 2.  Both of these spin-glass sponges exhibit surface spin-glass ordering in the absence of bulk spin-glass ordering.  The phase diagram has a surface-only spin-glass ordered phase, a surface-and-bulk spin-glass ordered phase, and a disordered phase.  The bulk ordering is not affected by the surface interaction, so that the phase boundary to the surface-and-bulk spin-glass ordered phase is a vertical line in the $J_SJ_B$ phase diagram, covering the phase transitions to both the surface-only spin-glass ordered phase and the disordered phase.  On the other hand, the bulk interaction $J_B$ enhances surface-only spin-glass ordering, even in the absence of bulk spin-glass ordering, as seen from the downwards deviation, as the bulk interaction $J_B$ increases, of the near-horizontal phase boundary between the surface-only spin-glass and disordered phases. As expected, the two phase boundaries for the same $d_S$ merge as $J_B$ goes to zero.  The three phase boundary lines meet at the special multicritical point M.  Sponge surfaces with $d_S \leq 2.46$ do not exhibit a surface-only spin-glass ordered phase.

\section{Distinctive Surface and Bulk Chaos: Lyapunov and Runaway Exponents}

The asymptotically chaotic renormalization-group trajectories starting within the surface or bulk spin-glass phase are shown in Fig. 3, where the consecutively renormalized (combining with neighboring interactions) values at a given location $<ij>$ are followed. The strength of chaos is measured by the Lyapunov exponent \cite{Collet,Hilborn}
\begin{equation}
\lambda = \lim _{n\rightarrow\infty} \frac{1}{n} \sum_{k=0}^{n-1} \ln \Big|\frac {dx_{k+1}}{dx_k}\Big|\,,
\end{equation}
where $x_k = J_{ij}/\overline{J}$ at step $k$ of the renormalization-group trajectory and $\overline{J}$ is the average of the absolute value of the interactions in the quenched random distribution.  We calculate the Lyapunov exponents by discarding the first 100 renormalization-group steps (to eliminate crossover from initial conditions to asymptotic behavior) and then using the next 900 steps.  The initial $(J_B,J_S)$ values do not matter, as long as they are within the specified spin-glass phase.  For each of our three systems, we calculated the Lyapunov exponents for sole surface chaos $\lambda_S$, driven surface chaos $\lambda_D$, and bulk chaos $\lambda_B$.  For the $b^{d_S}=6$, namely $d_S=2.63$ systems, after decimation the strong coupling interactions, in certain of the randomly chosen combinations, do nearly cancel each other, meaning frustration.  In these cases, the derivative in Eq. (2) is vanishingly small and not included in the summation, when the result of the bond-moving gives an interaction which is $10^{-6}$ times or less than the average interaction in the distribution.  Chaotic runaway does of course continue in the next renormalization-group iteration.

As seen in Fig. 3, in all three sponges, the Lyapunov exponent is larger, namely chaos is stronger for sole surface chaos than for driven chaos.  The Lyapunov exponent is even larger and chaos is even stronger for bulk chaos in all three systems.

In addition to chaos, the renormalization-group trajectories show asymptotic strong-coupling behavior,
\begin{equation}
\overline{J'} = b^{y_R}\, \overline{J}\,,
\end{equation}
where $y_R >0$ is the runaway exponent \cite{Demirtas}.  Again using 900 renormalization-group steps after discarding 100 steps, we calculated the runaway exponent for sole surface chaos $y_{RS}$, driven surface chaos $y_{RD}$, and bulk chaos $y_{RB}$.

\section{Conclusion}

We have studied surfaced spin-glass systems by exact renormalization-group theory of $d=3$ hierarchical lattices.  We find that a flat surface does not support chaotic rescaling and therefore does not have a surface spin-glass phase.  Therefore, we have constructed spin-glass sponges, with higher fractal dimensional surfaces.  We find that for higher spongevity, surface chaotic rescaling occurs even in absence bulk chaotic rescaling.  We calculate phase diagrams with sole surface spin-glass phase, surface-and-bulk spin-glass phase, and disordered phase, all three phases meeting at a special multicritical point $M$.  Chaos in the sole surface spin-glass phase is stronger than in the bulk-driven surface spin glass.

\begin{acknowledgments}
Support by the Kadir Has University Doctoral Studies Scholarship Fund and by the Academy of Sciences of Turkey (T\"UBA) is gratefully acknowledged.
\end{acknowledgments}

\end{document}